\documentclass[twocolumn,resetfootnote]{aastex701}

\usepackage{amsmath}

\begin{document}

\title{All-Sky Ultra-Narrowband Spectral Imaging with the OVRO–LWA: Technosignature Constraints and Axion-Like Particle Prospects}

\renewcommand{\sectionautorefname}{Section}

\newcommand{\affilone}{Cahill Center for Astronomy and Astrophysics, California Institute of Technology, Pasadena, CA 91125, USA}
\newcommand{\affiltwo}{Owens Valley Radio Observatory, California Institute of Technology, Big Pine, CA 93513, USA}
\newcommand{\affilthree}{Department of Physics and Astronomy, Rice University, Houston, TX 77005, USA}
\newcommand{\affilfour}{Jet Propulsion Laboratory, California Institute of Technology, Pasadena, CA 91011, USA}
\newcommand{\affilfive}{School of Earth and Space Exploration, Arizona State University, Tempe, AZ 85287, USA}
\newcommand{\affilsix}{Center for Solar-Terrestrial Research, New Jersey Institute of Technology, Newark, NJ 07102, USA}
\newcommand{\affilseven}{George Mason University, Fairfax, VA 22030, USA}
\newcommand{\affileight}{University of New Mexico, Albuquerque, NM 87131, USA}
\newcommand{\affilnine}{Real-Time Radio Systems Ltd, Bournemouth, Dorset, BH6 3LU, UK}
\newcommand{\affilelev}{Rice Space Institute, Rice University, Houston, TX 77005, USA}
\newcommand{\affiltwelve}{ASTRON, Netherlands Institute for Radio Astronomy, Oude Hoogeveensedijk 4, Dwingeloo, 7991 PD, The Netherlands}
\newcommand{\affilthirteen}{Department of Climate and Space Sciences and Engineering, University of Michigan, Ann Arbor, MI 48109, USA}
\newcommand{\affilfourteen}{National Centre for Radio Astrophysics, Tata Institute of Fundamental Research, Pune University Campus, Pune 411007, India}

\newcommand{\affilassign}[1]{%
    \ifnum#1=1 \affiliation{\affilone}\fi
    \ifnum#1=2 \affiliation{\affiltwo}\fi
    \ifnum#1=3 \affiliation{\affilthree}\fi
    \ifnum#1=4 \affiliation{\affilfour}\fi
    \ifnum#1=5 \affiliation{\affilfive}\fi
    \ifnum#1=6 \affiliation{\affilsix}\fi
    \ifnum#1=7 \affiliation{\affilseven}\fi
    \ifnum#1=8 \affiliation{\affileight}\fi
    \ifnum#1=9 \affiliation{\affilnine}\fi
    \ifnum#1=11 \affiliation{\affilelev}\fi
    \ifnum#1=12 \affiliation{\affiltwelve}\fi
    \ifnum#1=13 \affiliation{\affilthirteen}\fi
    \ifnum#1=14 \affiliation{\affilfourteen}\fi
}

\author[0000-0003-1226-118X]{Nikita Kosogorov}
\email[show]{nakosogorov@gmail.com} 
\email[show]{nkosogor@caltech.edu} 
\affilassign{1}
\affilassign{2}

\author[0000-0002-7083-4049]{Gregg Hallinan}
\email{gh@astro.caltech.edu}
\affilassign{1}
\affilassign{2}

\author[0000-0002-8191-3885]{Greg Hellbourg}
\email{ghellbourg@astro.caltech.edu}
\affilassign{1}
\affilassign{2}

\author[0000-0003-2238-2698]{Marin M. Anderson}
\email{marin.m.anderson@jpl.nasa.gov}
\affilassign{2}
\affilassign{4}

\author[0000-0002-8475-2036]{Judd D. Bowman}
\email{Judd.Bowman@asu.edu}
\affilassign{5}

\author[0000-0003-4980-2736]{Ruby Byrne}
\email{rbyrne@caltech.edu}
\affilassign{1}
\affilassign{2}

\author{Morgan Catha}
\email{morgan.catha@gmail.com}
\affilassign{2}

\author[0000-0002-0660-3350]{Bin Chen}
\email{bin.chen@njit.edu}
\affilassign{6}

\author[0000-0002-1810-6706]{Xingyao Chen}
\email{xingyao.chen@njit.edu}
\affilassign{6}

\author[0000-0001-7754-0804]{Sherry Chhabra}
\email{schhabr@gmu.edu}
\affilassign{7}

\author{Larry D’Addario}
\email{ldaddario@caltech.edu}
\affilassign{1}
\affilassign{2}

\author[0000-0001-5397-5969]{Ivey Davis}
\email{idavis@astro.caltech.edu}
\affilassign{1}
\affilassign{2}
\affilassign{12}

\author[0000-0003-1407-0141]{Jayce Dowell}
\email{jdowell@unm.edu}
\affilassign{8}

\author[0009-0007-6418-2498]{Katherine Elder}
\email{keelder@asu.edu}
\affilassign{5}

\author[0000-0003-2520-8396]{Dale Gary}
\email{dgary@njit.edu}
\affilassign{6}

\author{Charlie Harnach}
\email{charnach@caltech.edu}
\affilassign{2}

\author[0000-0003-0216-1417]{Jack Hickish}
\email{jackhickish@gmail.com}
\affilassign{9}

\author{Rick Hobbs}
\email{citrh@pm.me}
\affilassign{2}

\author{David Hodge}
\email{hodge@caltech.edu}
\affilassign{1}

\author{Mark Hodges}
\email{mwh@caltech.edu}
\affilassign{2}

\author[0000-0003-4267-6108]{Yuping Huang}
\email{yupinghyper@gmail.com}
\affilassign{1}
\affilassign{2}

\author[0000-0001-8061-2207]{Andrea Isella}
\email{ai14@rice.edu}
\affilassign{3}
\affilassign{11}

\author[0000-0002-0917-2269]{Daniel C. Jacobs}
\email{dcjacob2@asu.edu}
\affilassign{5}

\author{Ghislain Kemby}
\email{kemby@caltech.edu}
\affilassign{2}

\author{John T. Klinefelter}
\email{jtklinef@caltech.edu}
\affilassign{2}

\author[0000-0002-2950-2974]{Matthew Kolopanis}
\email{matthew.kolopanis@asu.edu}
\affilassign{5}

\author[0000-0002-5959-1285]{James Lamb}
\email{lamb@caltech.edu}
\affilassign{2}

\author[0000-0002-4119-9963]{Casey Law}
\email{claw@astro.caltech.edu}
\affilassign{1}
\affilassign{2}

\author[0000-0003-2560-8023]{Nivedita Mahesh}
\email{nmahesh@caltech.edu}
\affilassign{1}
\affilassign{2}

\author[0000-0002-2325-5298]{Surajit Mondal}
\email{surajit.mondal@njit.edu}
\affilassign{14}

\author[0000-0003-2802-850X]{Brian O’Donnell}
\email{beo6@njit.edu}
\affilassign{6}

\author[0000-0001-6360-6972]{Kathryn Plant}
\email{kathryn.a.plant@jpl.nasa.gov}
\affilassign{2}
\affilassign{4}

\author{Corey Posner}
\email{cposner@caltech.edu}
\affilassign{2}

\author{Travis Powell}
\email{tpowellky14@gmail.com}
\affilassign{2}

\author{Vinand Prayag}
\email{vinand@caltech.edu}
\affilassign{2}

\author{Andres Rizo}
\email{rizoa@caltech.edu}
\affilassign{2}

\author[0000-0002-4992-4162]{Andrew Romero-Wolf}
\email{andrewrw@hawaii.edu}
\affilassign{4}

\author[0000-0003-1647-7762]{Jun Shi}
\email{shijun88vip@163.com}
\affilassign{1}

\author[0000-0001-6495-7731]{Greg Taylor}
\email{gbtaylor@unm.edu}
\affilassign{8}

\author{Jordan Trim}
\email{jtrim@caltech.edu}
\affilassign{2}

\author{Mike Virgin}
\email{mvirgin@caltech.edu}
\affilassign{2}

\author[0000-0002-6611-2668]{Akshatha Vydula}
\email{vydula@asu.edu}
\affilassign{5}

\author[0000-0002-9353-6204]{Sandy Weinreb}
\email{sweinreb@caltech.edu}
\affilassign{1}

\author{Scott White}
\email{swhite2036@gmail.com}
\affilassign{2}

\author[0000-0003-1734-2472]{David Woody}
\email{dpw@ovro.caltech.edu}
\affilassign{2}

\author[0000-0003-2872-2614]{Sijie Yu}
\email{sijie.yu@njit.edu}
\affilassign{6}

\author{Thomas Zentmeyer}
\email{thomas.zent@gmail.com}
\affilassign{2}

\author[0000-0001-6855-5799]{Peijin Zhang}
\email{peijin.zhang@njit.edu}
\affilassign{6}

\author{T. Joseph W. Lazio}
\email{lazio@caltech.edu}
\affilassign{13}


\begin{abstract}
We present an imaging-domain search for technosignatures at decametric wavelengths with the OVRO--LWA, targeting ultra-narrowband continuous-wave signals between 50 and 86\,MHz. We implement an offline GPU pipeline that processes raw voltage data with upchannelization to approximately 10\,Hz frequency resolution, producing all-sky images for each fine channel and totaling more than $3\times10^{6}$ images for a single 30\,s epoch. Candidate selection is performed using multi-kernel matched filtering across frequency, empirical noise standardization, and false-discovery-rate control. After applying quality cuts that remove extended sources, corrupted images, and obvious RFI, three narrowband candidates with signal-to-noise ratios above $10\sigma$ were selected for detailed analysis. By re-imaging these candidates with finer temporal and spectral resolution, we resolved their structure and found them to be inconsistent with compact celestial narrowband emitters. Consequently, we report no detection of extraterrestrial technosignatures. The representative sensitivity of the search is $\sim\!100$\,Jy per channel across the entire visible hemisphere. For an unresolved emitter, this corresponds to $10\sigma$ equivalent isotropic radiated power (EIRP) limits of about $10^{14}$\,W at a distance of 10\,pc and $10^{18}$\,W at 1\,kpc. The wide field of view and ultra-fine spectral resolution of this approach enable simultaneous probing of technosignature signals from millions of stellar systems. This method further establishes a scalable framework for deeper integrations and stacked searches toward neutron-star targets relevant to axion-like particle (ALP) line conversion.
\end{abstract}



\section{Introduction}
The search for radio technosignatures—the engineered emissions of technological civilizations—remains one of the most mature and observationally tractable approaches to addressing the question ``Are we alone in the Universe?'' Conducted across meter- to centimeter-wavelengths, these experiments exploit radio astronomy's fine spectral resolution, wide instantaneous bandwidths, and the transparency of the interstellar medium over much of the radio band. Foundational arguments for interstellar radio communication and the ``watering hole'' near the H\,\textsc{i} line framed the early searches \citep[e.g.,][]{1959Natur.184..844C,1961PhT....14d..40D,2020RAA....20...78L,2001ARA&A..39..511T,2018AJ....156..260W,2018arXiv181208681N}. Modern programs—e.g., Breakthrough Listen’s multi-GHz campaigns with the Robert C. Byrd Green Bank Telescope (GBT), Murriyang (Parkes Radio Telescope), the Five-hundred-meter Aperture Spherical Telescope (FAST), and the Meer Karoo Array Telescope (MeerKAT) pathfinders—refine drift-search pipelines and implement rigorous on/off, multibeam discrimination \citep[e.g.][]{2017AcAau.139...98W,2017ApJ...849..104E,2020AJ....159...86P,2021PASP..133f4502C,2021PhDT.........5S,2024AJ....168..284B,2023AJ....166..206M,2024RASTI...3...33M}, while extending coverage across increasingly broad portions of the radio spectrum, though still predominantly at decimeter- to centimeter wavelengths ($\sim 0.6$--8\,GHz).

In recent years, a growing fraction of the search space \citep{2020IJAsB..19..237S} has opened at low radio frequencies ($\lesssim 300$\,MHz). These bands, historically underexplored relative to the GHz regime, offer unique advantages: lower-cost high-equivalent isotropic radiated power (EIRP) beacons, leakage analogs of terrestrial broadcast/aviation/maritime services, enormous instantaneous fields of view, and excellent survey speed with sparse-aperture arrays. Recent dual-station Low-Frequency Array (LOFAR) experiments have demonstrated the effectiveness of simultaneous multi-site anti-coincidence RFI rejection across $\sim10^6$ TESS/Gaia targets in the 110--190\,MHz band \citep{2023AJ....166..193J}, extending earlier anti-coincidence strategies developed in radio SETI \citep{1992AcAau..26..219W,2005RaSc...40.5S18H,2016AJ....152..181H} into the low-frequency regime.

Low-frequency, ultra-fine spectral imaging is also directly applicable to ultralight dark matter searches. Axions arise as pseudo-Nambu–Goldstone bosons from the Peccei–Quinn solution to the strong-CP problem, while more general axion-like particles (ALPs) share the two-photon coupling but are not tied to quantum chromodynamics (QCD); both can mix with photons in magnetized plasmas \citep[e.g.,][]{2015ARNPS..65..485G,2016PhR...643....1M,2018PrPNP.102...89I,2021JCAP...11..013M,2022SciA....8J3618C}. In neutron-star magnetospheres, axion/ALP dark matter can resonantly convert into narrow radio lines where the local plasma frequency matches the particle mass, yielding an observed line near
\begin{equation}
\nu \simeq 242~\mathrm{MHz}\,\left(\frac{m_a}{\mu\mathrm{eV}}\right),
\end{equation}
where $\nu$ is the photon frequency of the converted radio line and $m_a$ is the axion/ALP mass expressed in energy units. This motivates targeted radio searches toward magnetars and isolated neutron stars \citep[e.g.,][]{2018PhRvL.121x1102H,2019PhRvD..99l3021S,2020PhRvD.101l3003L,2021JHEP...09..105B,2021PhRvD.104j3030W}. The intrinsic line width is primarily set by the dark matter velocity dispersion, typically $\Delta\nu / \nu \sim 10^{-6}$, placing such signals well within the regime probed by ultra-narrow spectral channels.

Detectability at tens of megahertz is strongly model-dependent: at these low frequencies the resonant layer lies farther from the neutron star, where magnetic fields are weaker and propagation effects can diminish the escaping flux. Even so, wide-field low-frequency arrays remain compelling for ALPs with enhanced photon couplings and for population-level searches that stack many neutron stars statistically. Analogous conversion processes have also been explored for other light bosons, such as dark photons, with recent LOFAR observations of the solar corona between $\sim$30–80\,MHz placing new constraints on dark-photon mixing \citep{2024NatCo..15..915A}.

Lastly, a complementary channel is two-photon decay $a\!\to\!\gamma\gamma$: while spontaneous decay is negligible for $\mu$eV axions, a bright ambient radio field can yield large stimulated emission, motivating diffuse, narrow-line searches toward bright backgrounds (e.g., the inner Galaxy) \citep{2019JCAP...03..027C,2024JCAP...04..045B}.

Both science cases therefore motivate instruments that combine ultra-fine spectral resolution with wide-field radio imaging that can be implemented efficiently at large data volumes. Meeting these requirements increasingly relies on software-defined interferometers, in which digital channelization, correlation or beamforming, radio-frequency interference (RFI) excision, and imaging are implemented on graphical-processing-unit (GPU) clusters. Examples include LOFAR and its ongoing LOFAR2.0 upgrade, which combines distributed clock synchronization with software-based correlation and beamforming running on large compute clusters, enabling simultaneous low- and high-band operation and flexible calibration and survey modes \citep{2013A&A...556A...2V, 2021A&A...652A..37E,2024SPIE13096E..2GH}. The Murchison Widefield Array (MWA) Phase II combined a reconfigurable layout with a GPU-based correlator and real-time calibration and imaging pipeline, improving $uv$-coverage, angular resolution, and wide-field imaging sensitivity \citep{2018PASA...35...33W}. On a general-purpose dish array, the Karl G. Jansky Very Large Array (VLA) has been equipped with the Commensal Open-Source Multimode Interferometer Cluster (COSMIC) backend, a GPU-based pipeline running during routine operations for technosignature and narrow-line searches \citep{2024AJ....167...35T}. One of the upcoming facilities, the Deep Synoptic Array (DSA), is explicitly designed as a radio camera, producing science-ready images nearly in real time and exemplifying the software-telescope paradigm \citep{2019BAAS...51g.255H,2024AAS...24323705H}. Within this evolving landscape, the Owens Valley Radio Observatory Long Wavelength Array (OVRO–LWA) (G. Hallinan et al., in prep) brings these capabilities to the 13–86\,MHz regime, enabling all-sky, high-cadence imaging ideally suited for broadband transient and narrowband line searches, while also serving as a pathfinder for DSA.

The OVRO--LWA is a low-frequency interferometer located at Owens Valley Radio Observatory, California ($37.24^\circ$ N, $118.28^\circ$ W). While using the same basic dipole design as the New Mexico LWA stations \citep[e.g.,][]{2012JAI.....150004T}, it was independently engineered for all-sky, arcminute-scale imaging, requiring extensive custom hardware. The array was recently expanded to 352 dual-polarization antennas: 241 in a $\sim$200\,m core and 111 in an extended layout with baselines up to 2.4\,km. The system covers 13–86.5\,MHz with native 24\,kHz channels, cross-correlating all antennas to deliver $\sim$20,000\,deg$^2$ wide-field images at resolutions from $\sim$5 to 30 arcmin across the band. The array facilitates a diverse range of science cases, covering the search for exoplanetary magnetospheres and the tracking of stellar space weather \citep[e.g.,][]{2025ApJ...993...82D}, as well as probing the Cosmic Dawn via the 21\,cm hydrogen line \citep[e.g.,][]{2018AJ....156...32E,2019AJ....158...84E,2021MNRAS.506.5802G}. Further capabilities include dynamic spectroscopic imaging of solar phenomena \citep[e.g.,][]{2021ApJ...906..132C,2025ApJ...992..143M,2025ApJ...992..128Z,2025ApJ...990L..50C,2025ApJ...994..254M,2026ApJ..1003...57Z}, measuring cosmic ray air showers \citep[e.g.,][]{2020NIMPA.95363086M,2022icrc.confE.204P,2026arXiv260313205P}, monitoring for low-frequency transient events \citep[e.g.,][]{2018ApJ...864...22A,2019ApJ...886..123A}, analyzing radio afterglows from meteor trails \citep[e.g.,][]{2024JGRA..12932272V}, and studying galaxy clusters \citep[e.g.,][]{2023MNRAS.521.5786H}.

More recently, a system was developed at the OVRO--LWA specifically for fast-transient searches of prompt radio emission from gravitational wave (GW) events, with its full design, validation and application to a real GW event described in \citet{2025ApJ...985..265K} and \citet{2026ApJ...997..311K}. This system introduced upchannelized beamforming and imaging, allowing for extremely narrow frequency channels of about 700\,Hz. In this work, we build on this capability for ultra-narrow-channel imaging at about 10\,Hz resolution.

In this paper, we explore how the OVRO--LWA can be applied to searches for technosignatures and ALPs. In \autoref{sec:obs-prop}, we present the observations, data processing, propagation and kinematic considerations relevant to narrowband signals, and the procedures used for candidate identification, vetting, and higher-resolution follow-up. In \autoref{sec:analysis}, we translate the search into technosignature constraints and discuss the implications for ALP searches. \autoref{sec:summary} summarizes the main results and future directions.

\begin{figure*}[htbp]
  \centering
  \includegraphics[width=\textwidth]{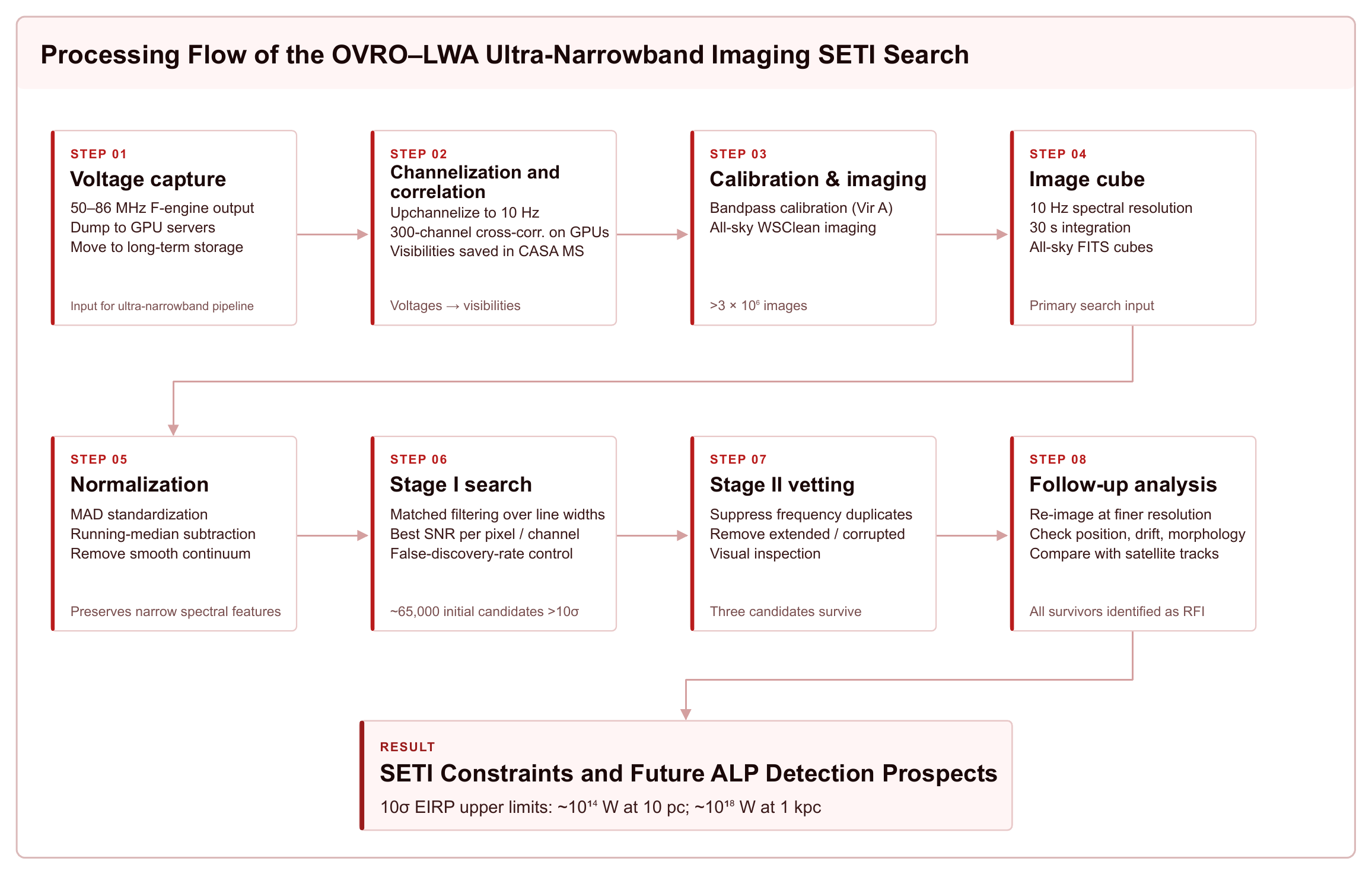}
  \caption{Summary of the OVRO--LWA ultra-narrowband imaging search pipeline. Raw voltage data are upchannelized and cross-correlated (\autoref{sec:xcorr}), calibrated and imaged (\autoref{sec:calib}) into all-sky 10\,Hz, 30\,s image cubes. These are then normalized and searched for candidates (\autoref{sec:cands}), followed by vetting and higher-resolution analysis (\autoref{sec:cand_analysis}, \autoref{sec:high-res}). The results are presented in \autoref{sec:analysis}.}
  \label{fig:flowchart}
\end{figure*}

\begin{figure*}[htbp]
  \centering
  \includegraphics[width=\textwidth]{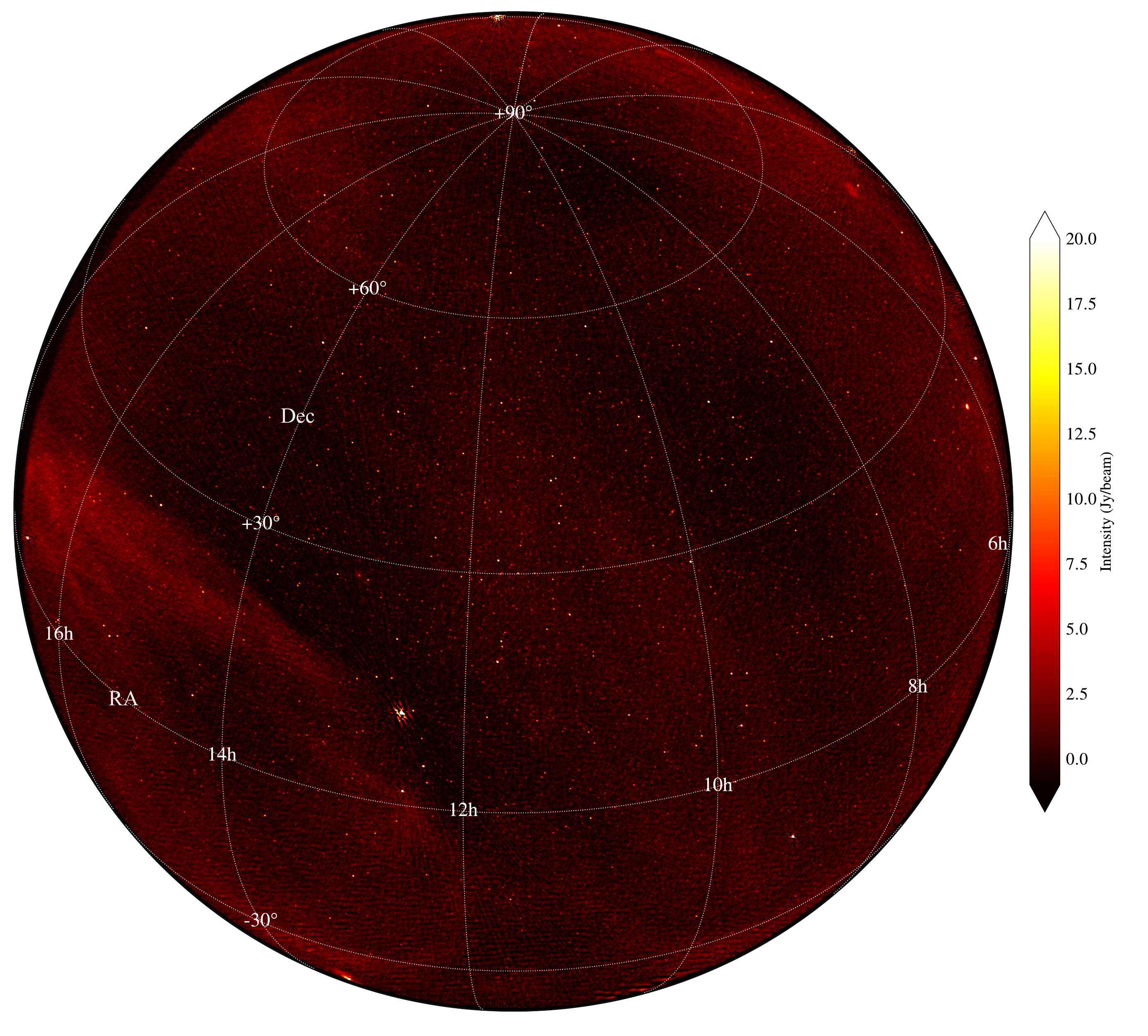}
  \caption{All-sky OVRO--LWA map in J2000 equatorial coordinates, generated from real-time cross-correlation data for a single 10\,s integration across the 50--86\,MHz band. Right ascension is shown in hours and declination in degrees. The synthesized beam is $549.4\arcsec \times 453.7\arcsec$ (FWHM) at a position angle of $44.9^\circ$.}

  \label{fig:allsky_all_chan}
\end{figure*}

\section{Observations, Data Processing and Signal Detectability}\label{sec:obs-prop}
This section describes the OVRO–LWA observations and data-processing pipelines used to produce ultra-narrowband, all-sky images, followed by a summary of propagation and kinematic effects that inform the detectability and interpretation of narrowband signals, and the subsequent procedures for candidate detection and validation. Figure~\ref{fig:flowchart} provides an overview of the full processing pipeline used in this work.

\begin{figure*}[htbp]
  \centering
  \includegraphics[width=\textwidth]{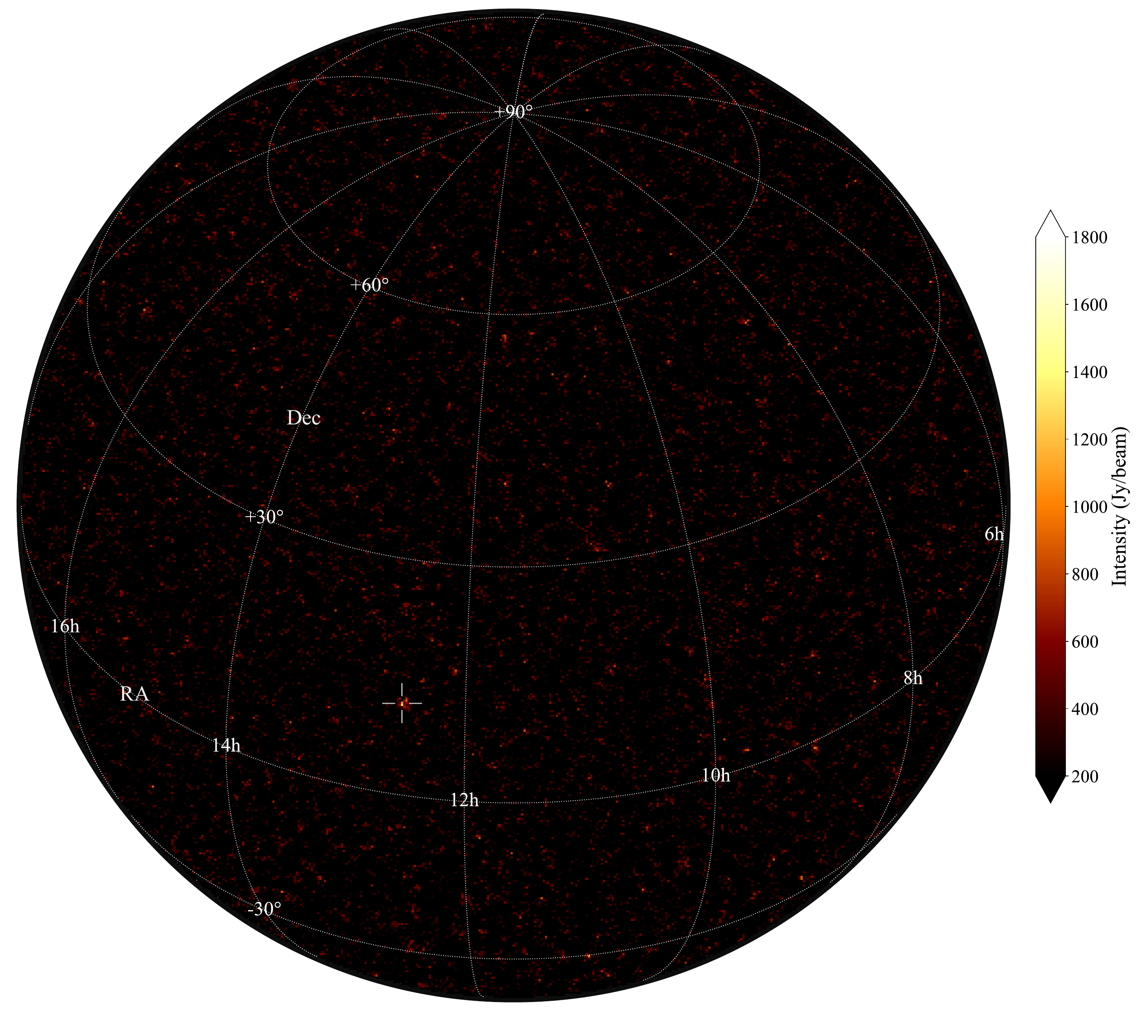}
  \caption{All-sky OVRO--LWA ultra-narrowband ($\sim$10\,Hz) map in J2000 equatorial coordinates, generated from the offline cross-correlation pipeline for a 30\,s integration at 72,675,248 Hz. Right ascension is shown in hours and declination in degrees. The synthesized beam is $41.2\arcmin \times 25.2\arcmin$ (FWHM) at a position angle of $125.5^\circ$. Because this image corresponds to a single 10\,Hz channel, its noise is much higher than in the wideband image, and most broadband continuum emission and most point sources are not significantly detected. Vir~A, near RA $= 12^{\mathrm h}30^{\mathrm m}$ and Dec $= +12.4^\circ$, is clearly visible; the white cross marks its position.}

  \label{fig:allsky_one_chan}
\end{figure*}

\subsection{Cross-Correlation Pipeline}\label{sec:xcorr}
For a detailed description of the OVRO–LWA data buffer dump and offline cross-correlation pipeline, see \citet{2025ApJ...985..265K}. Below, we summarize the main components relevant to our analysis and the modifications implemented for this work.

We first record the raw output of the OVRO--LWA F-engine after polyphase filter bank (PFB) processing across the 50--86\,MHz band and dump it to the GPU servers for initial storage. These data are then transferred to our 5\,PB storage system for further processing. We then use the offline cross-correlation pipeline, which upchannelizes the data by applying a fast Fourier transform (FFT) to the time samples within each coarse channel, thereby increasing the spectral resolution. For fast-transient searches associated with GW events, the typical upchannelization factor is 32. Using the default online channelization of about 23.9\,kHz, this results in channels of about 750\,Hz. In our case, however, we choose a factor of 2400, resulting in a final channel resolution of 9.9\,Hz. Given the large data volume produced by upchannelization, we divide the data into manageable chunks of 300 fine channels, corresponding to approximately 3\,kHz of bandwidth per batch. For each batch, the selected fine channels are cross-correlated on a GPU and the procedure is then repeated for the next batch of 300 fine channels until the full 50--86\,MHz band is covered. Independent batches can be distributed across the available GPUs, allowing multiple portions of the band to be processed in parallel while keeping the memory footprint of each GPU job manageable.

Cross-correlation is implemented within the Bifrost framework \citep{2017JAI.....650007C} and executed on available GPUs using dedicated processing blocks\footnote{\url{https://github.com/realtimeradio/caltech-bifrost-dsp}}. The correlation stage is accelerated by a GPU kernel built on the xGPU library\footnote{\href{https://github.com/GPU-correlators/xGPU}{https://github.com/GPU-correlators/xGPU}} \citep[][]{2011arXiv1107.4264C}. The resulting visibilities are stored in Hierarchical Data Format (HDF) files with an integration time of 30\,s. Subsequently, the visibilities are converted into the Common Astronomy Software Applications (CASA)\footnote{\url{https://casa.nrao.edu/}} \citep[][]{2022PASP..134k4501C} Measurement Set (MS) format while retaining all necessary metadata.  

The dataset analyzed in this work was obtained on 2024-04-06 at 06:30:57 UTC and consists of a single 30\,s observing epoch used for the primary search. For candidate follow-up, we additionally reprocessed subsequent 30\,s integrations spanning approximately three minutes. Although the Galactic plane dominates the sky temperature at tens of MHz, its emission is spectrally smooth. Consequently, while the Galaxy contributes to the overall system noise, it does not mimic the spectral characteristics of ultra-narrowband features. We did not impose any constraint on the Galactic-plane elevation, as our search algorithm effectively separates this broadband continuum from narrow spectral lines.
The current configuration of 10\,Hz channelization and 30\,s integration is motivated by the need to probe fractional linewidths of order $\Delta\nu/\nu \sim 10^{-6}$ at tens of MHz, while remaining computationally feasible. For this analysis, we utilized eight NVIDIA RTX A4000 GPUs to perform the offline cross-correlation.

\subsection{Calibration and Imaging Pipeline }\label{sec:calib}

We then proceeded with calibrating the resulting visibility dataset. For gain calibration, we adopted a simplified sky model consisting of flux densities and spectral indices for the brightest radio sources \citep[][]{1977A&A....61...99B,2017ApJS..230....7P}. At the time of observation, the only sufficiently bright calibrator above the horizon was Virgo~A. The same measurement sets were then used to calibrate each channel individually, as the calibrator was bright enough for this purpose.

The calibration included a bandpass correction using the \texttt{bandpass} task within the Common Astronomy Software Applications (CASA)\footnote{\url{https://casa.nrao.edu/}} \citep{2022PASP..134k4501C}. Imaging was performed with \texttt{WSClean}\footnote{\url{https://gitlab.com/aroffringa/wsclean}} \citep{2014MNRAS.444..606O} over the visible hemisphere on a $512 \times 512$ grid.

We show the online cross-correlated data image in \autoref{fig:allsky_all_chan} for the same Local Sidereal Time (LST), and in \autoref{fig:allsky_one_chan} we present an example 10\,Hz, 30\,s integration image of the full sky. The wideband reference image shown in \autoref{fig:allsky_all_chan} was made with a finer imaging grid, whereas the ultra-narrowband search images, including \autoref{fig:allsky_one_chan}, were produced on the coarser $512\times512$ grid with $0.25^\circ$ pixels. This coarser grid was adopted to make the production and storage of more than $3\times10^6$ images tractable, and the larger restoring beam in \autoref{fig:allsky_one_chan} reflects the effective resolution of this search-image product. These images are retained as the sole data products for subsequent analysis. The $10~\mathrm{Hz}$, $30~\mathrm{s}$ image has a sensitivity of about $120~\mathrm{Jy}$ (1$\sigma$). The Virgo~A detection shows a signal-to-noise ratio (SNR) of $\sim$19, consistent with the expected value of $\sim$20, indicating that the image noise and flux calibration are within expected limits.

\subsection{Propagation and Kinematic Effects}\label{sec:prop}
Before introducing candidate selection, we summarize the propagation and kinematic effects that shape intrinsically narrow tones in our images. These effects determine the apparent width, amplitude modulation, and detectability of any true narrowband emitter.

\subsubsection{Dispersion and Scintillation}\label{sec:scint}
Cold-plasma dispersion introduces a deterministic, frequency-dependent group delay that affects all signals propagating through the interstellar medium. For a steady, continuous narrowband tone, this delay is constant in time and therefore has no observable effect on detectability. Within a 10\,Hz channel at our observing frequencies, the intra-channel delay predicted by the standard dispersion relation is below the millisecond level for ${\rm DM}\sim1$--$10\,{\rm pc\,cm^{-3}}$, which is negligible compared to our 30\,s integrations. Dispersion is thus only relevant for transient signals on sub-second timescales.

To assess diffractive interstellar scintillation (DISS), we convert empirical pulse-broadening timescales to equivalent diffractive decorrelation bandwidths using standard thin-screen relations. We adopt the pulse-broadening--DM scalings, together with their large intrinsic sightline-to-sightline scatter, as
characterized by both \citet{2004ApJ...605..759B} and the updated analysis of \citet{2022ApJ...931...88C}. At our observing frequencies and ${\rm DM}\sim$ a few--tens of ${\rm pc\,cm^{-3}}$, the typical diffractive bandwidths are tens of Hz to kHz, while the observed scatter allows substantially smaller values, extending to $\lesssim$\,Hz along strongly scattered lines of sight. Consequently, the number of independent scintles across a 10\,Hz channel can range from order unity to $\gg 1$, depending on the line of sight.

The DISS timescale, set by the drift of the diffraction pattern across the line of sight (and hence by the effective transverse velocity), is typically of order minutes at $\sim$GHz frequencies \citep[e.g.][]{1990ARA&A..28..561R,1998ApJ...507..846C,2019MNRAS.485.4389R}. 
For a Kolmogorov medium, the diffractive scintillation timescale scales with observing frequency as $\Delta t_{\rm d}\propto \nu^{6/5}$ \citep[e.g.,][]{2018ApJ...861...12L}. Applying this scaling to meter wavelengths implies characteristic DISS timescales of order seconds to tens of seconds across our observing band, with $\sim$10--30~s as a representative value. Over a 30~s integration, this corresponds to sampling roughly one to a few independent scintles in time, such that the resulting modulation index can range from near unity down to a few tenths, depending on the line of sight.

In summary, the total intensity modulation can vary widely, from about 100\% down to below 10\%, depending on both the diffractive bandwidth and the temporal scintillation rate along a given sightline.

\subsubsection{Free–Free Absorption and Angular Broadening}\label{sec:absorp}

We estimate thermal free–free optical depth using the \citet{1967ApJ...147..471M} approximation with a representative electron temperature of about 8000\,K. At a representative frequency of 70\,MHz, this yields optical depth $\tau \approx 1.2\times 10^{-4}$ per ${\rm pc,cm^{-6}}$ of emission measure (EM), with only factor-of-unity variation across the 50--86\,MHz range. For smooth, low-DM paths off the Galactic plane, the implied EMs are small and $\tau \ll 10^{-3}$. Even for much denser, clumpier paths with ${\rm EM}\sim 100\,{\rm pc\,cm^{-6}}$, attenuation remains at the percent level. We therefore neglect free–free absorption for most sightlines considered here.

For a Kolmogorov spectrum of density fluctuations, image broadening steeply increases toward lower frequency. Empirically, off-plane sightlines at these frequencies show sub-arcsecond broadening, while generic paths in the plane show arcsecond-scale values. Only the strongest inner-Galaxy directions, such as toward the Galactic Center, reach arcminute scales at these frequencies. With a 5--10 arcmin synthesized beam, angular broadening is negligible for the vast majority of the sky and only potentially comparable along the most strongly scattered lines of sight.

\subsubsection{Spectral broadening from intervening plasma}\label{sec:exoipm}
Propagation through circumstellar or interplanetary plasma may broaden an intrinsically ultra-narrow technosignature, especially at low observing frequencies and for lines of sight passing close to the host star; transient activity such as coronal mass ejections may enhance this effect further \citep{2026ApJ...999..210G}. Interstellar scattering can likewise broaden intrinsically narrow signals and, in principle, may impose an effective horizon once the broadened width becomes comparable to or larger than the search channel width \citep[e.g.,][]{1991ApJ...376..123C,1997ApJ...487..782C}. Such broadening is expected to be more pronounced at low observing frequencies and along strongly scattered sightlines \citep[e.g.,][]{2022PASA...39....8T}. We do not model this source-by-source here, but it motivates searching not only for single-channel excesses but also for modestly broadened features. A more detailed sightline-dependent treatment using modern Galactic electron-density models, such as NE2025, is left for future work \citep{2026ApJ..1002....3O}.

\subsubsection{Doppler drift}

Doppler drift is a kinematic effect arising from the relative acceleration of the transmitter and receiver. For the receiver alone (Earth's rotation and orbital motion), the expected drift at 70\,MHz over a 30\,s integration is small compared to our 10\,Hz channel width. Scaling the standard narrowband SETI estimates for Earth-induced drift rates to 70\,MHz gives a maximum receiver-induced drift of only a few tenths of a Hz over 30\,s, even in the most unfavorable line-of-sight case \citep{2019ApJ...884...14S,2020AJ....159...86P}. Thus, for most signals the power remains concentrated in the same 10\,Hz channel during a single integration.

A signal that begins very near a channel boundary can still drift into an adjacent bin, but for receiver-induced drift alone this should affect only a small minority of cases, typically at the few-percent level. In this sense, Doppler drift from Earth's motion is usually a minor correction for our present 30\,s, 10\,Hz analysis.

For transmitters bound to planets, rotational and orbital accelerations can dominate the observed drift \citep{2019ApJ...884...14S}. In particular, close-in giant planets around Sun-like stars can plausibly produce total drifts of tens of Hz over 30\,s, depending on geometry, while tidally locked rotation typically contributes a smaller but still non-negligible term. Our current search does not apply drift-rate corrections; future implementations can include a modest drift search spanning at least tens of Hz over 30\,s to capture the most extreme close-in planetary cases \citep{2019ApJ...884...14S}.

\subsection{Candidate Identification Pipeline}\label{sec:cands}

We use a two-stage procedure to identify and initially vet candidates. In \emph{Stage~I}, we search the standardized image cube for statistically significant narrowband outliers over a range of assumed linewidths. In \emph{Stage~II}, we reject duplicate detections arising from adjacent frequency bins or multiple trial linewidths at the same sky position, along with corrupted images and candidates inconsistent with unresolved narrowband point sources.

We first standardize each channelized all-sky image by estimating the median absolute deviation (MAD) and converting the result to maps with approximately unit variance. To remove broad spectral structure, we subtract a running median along the frequency axis at each sky pixel, using a window much wider (over 5~kHz) than the widest lines in our search (about 1~kHz). This high-pass step preserves narrowband features while suppressing continuum emission, bandpass ripples, and other slowly varying spectral trends. 

\begin{figure}[htbp]
  \centering
  \includegraphics[scale=0.5]{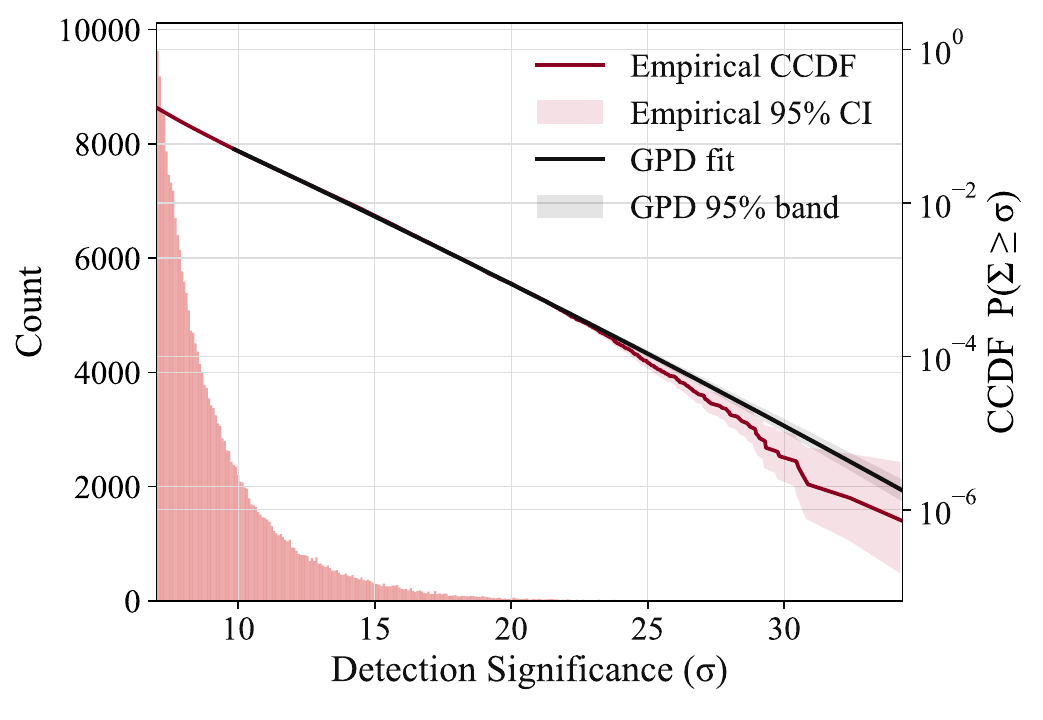}
  \caption{Histogram of the initially identified \emph{Stage~I} candidates as a function of SNR (left axis). The corresponding empirical complementary cumulative distribution function (CCDF; right axis) is shown together with a fitted generalized Pareto distribution (GPD; solid line) and bootstrap confidence interval (shaded band).
}

  \label{fig:ccdf_spatial}
\end{figure}

\begin{figure}[htbp]
  \centering
  \includegraphics[scale=0.75]{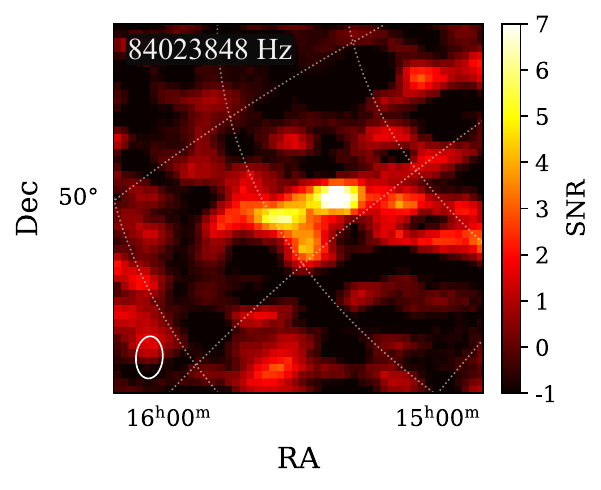}
  \caption{Example of a narrowband detection rejected during \emph{Stage~II} because its morphology is clearly extended relative to the synthesized beam. The ellipse in the lower-left corner shows the synthesized beam FWHM.}
  \label{fig:extended_reject}
\end{figure} 

To accommodate uncertain intrinsic linewidths, we convolve each per-pixel frequency series with a set of unit-normalized boxcar kernels spanning widths from 1 to 100 fine channels. For our $\simeq10$\,Hz channel spacing, this corresponds to linewidths from $\sim$10\,Hz to $\sim$1\,kHz. The kernel widths are sampled densely at the narrowest linewidths, with single-channel ($\simeq10$\,Hz) spacing up to around $\sim$100\,Hz, followed by progressively coarser spacing at broader linewidths: steps of a few fine channels over $\sim$100--500\,Hz and tens of fine channels near the upper end of the $\sim$1\,kHz range. This width-aware search is also motivated by the possibility that intrinsically ultra-narrow extraterrestrial signals may be spectrally broadened during propagation through circumstellar plasma, especially at low radio frequencies \citep{2026ApJ...999..210G}. For every pixel and channel, we retain the highest SNR and record the corresponding effective width. This approach enhances sensitivity to slightly broadened lines without restricting the search to a single linewidth.

\begin{figure*}[htbp]
  \centering
  \includegraphics[width=0.9\textwidth]{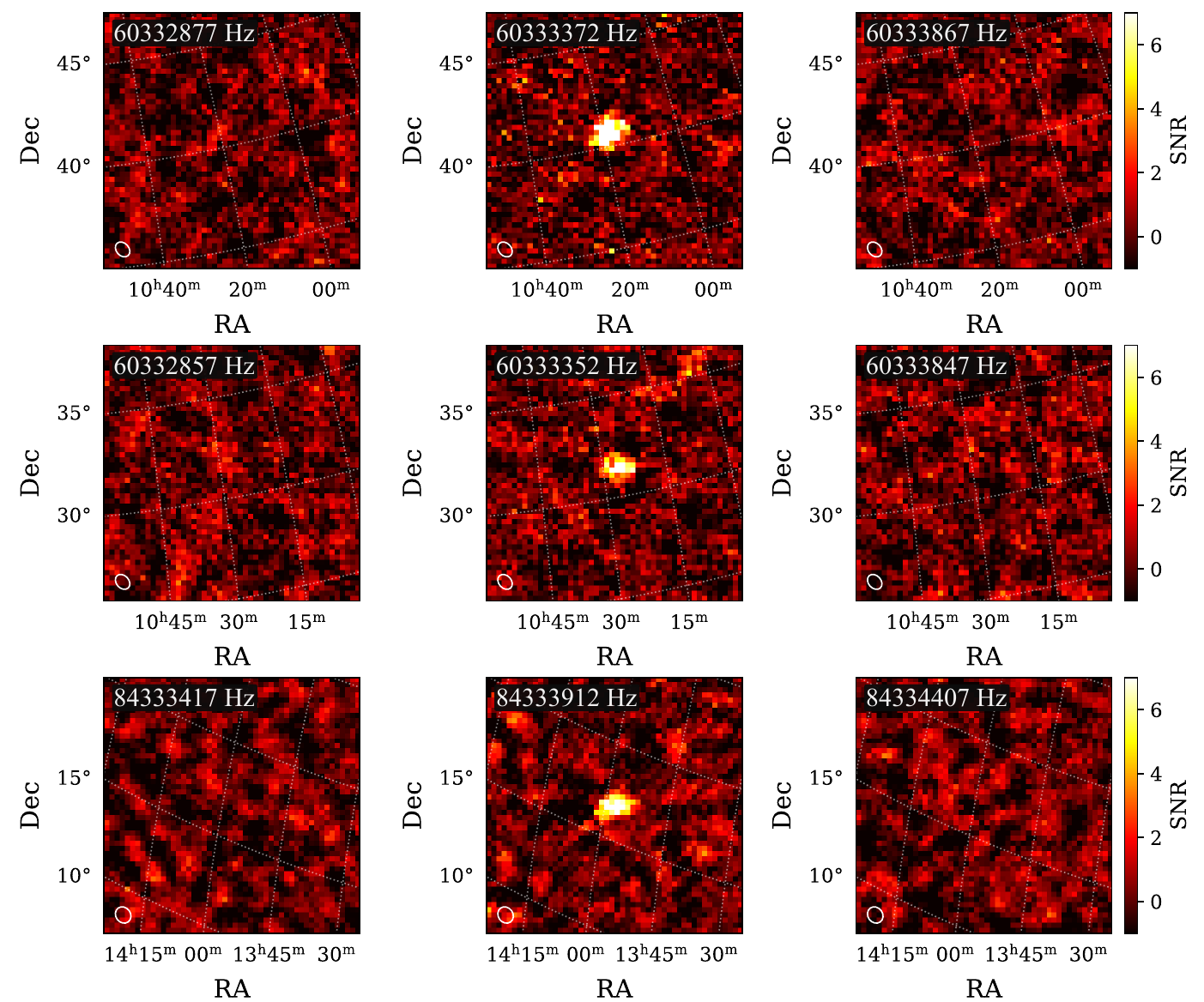}
  \caption{Summary of the three surviving narrowband candidates after \emph{Stage~II} vetting. Their corresponding parameters are presented in \autoref{tab:stage2_candidates}. The left and right panels show example images at frequencies offset by $-50$\,channels ($\sim$$-500$\,Hz) and $+50$\,channels ($\sim$$+500$\,Hz) from the detected frequency, respectively, while the middle panel shows the image at the detected frequency. These offsets were chosen as representative nearby comparisons; other nearby off-line channels show the same qualitative absence of emission. The ellipse in the lower-left corner shows the synthesized beam FWHM.}
  \label{fig:all_candidates}
\end{figure*}

Instead of applying a fixed SNR threshold, we convert SNR values to $p$-values under the standardized noise model and control the false discovery rate (FDR) using the Benjamini--Hochberg procedure \citep{Benjamini1995} across all searched pixels and channels. This yields an adaptive detection threshold with explicit control of the expected false-positive fraction across the full search. Neighboring image pixels are not strictly independent because of the synthesized beam, and adjacent spectral channels acquire weak correlations from the spectral filtering and baseline subtraction; therefore, the corresponding false-positive rate is approximate.

Because the search is performed with multiple spectral kernels, a single underlying narrowband feature can generate several nearby detections at the same sky pixel, either in adjacent frequency bins or for multiple trial widths. To collapse such repeated triggers, we apply one-dimensional non-maximum suppression along the frequency axis independently at each sky pixel. 
For each provisional detection, we define a local exclusion region centered on its detection frequency with half-width equal to half of the matched-filter width, rounded down to the nearest channel, with a minimum half-width of one channel. We then retain only the highest-SNR detection within this region. This step merges multiple nearby detections arising from the same spectral feature into a single representative candidate. Each surviving detection is then re-evaluated within a local spectral window by re-estimating the small-scale baseline and recomputing the SNR; detections that fail to reproduce at or above the decision threshold are discarded.

For each surviving candidate, we record the sky position, detection frequency, effective linewidth, and SNR. \autoref{fig:ccdf_spatial} summarizes the distribution of all initially detected candidates. The total number of initial \emph{Stage~I} candidates with matched-filter SNR above 10 is approximately~65{,}000. We refer to this as \emph{Stage~I} of our candidate-identification pipeline. Although the initial detection step is defined through Benjamini--Hochberg-controlled $p$-values, we quote matched-filter SNR values here because they are more directly interpretable when summarizing candidate properties. These detections are dominated by RFI, image corruptions, and residual artifacts, and are strongly concentrated in a relatively small number of image channels and narrow spectral regions, including low-VHF portions of the band associated with heavily used terrestrial allocations such as legacy television low-band and adjacent fixed/mobile-control services.

The next stage of processing, which we call \emph{Stage~II}, focuses on identifying candidates consistent with narrowband point sources. We automate this process by rejecting detections with extended morphology (see, e.g., \autoref{fig:extended_reject}) and by flagging images whose noise properties deviate significantly from expectations. These automated cuts reduced the candidate list to 12 detections with SNR above our conservative threshold of 10. Visual inspection of these remaining detections rejected nine additional artifacts or obvious RFI-like cases, leaving three apparently compact candidates for detailed follow-up. \autoref{tab:stage2_candidates} below lists the corresponding parameters, and we show them in \autoref{fig:all_candidates}. The sky coordinates reported in \autoref{tab:stage2_candidates} are obtained from Gaussian centroid fits to the candidate images. The quoted $1\sigma$ uncertainties are derived from the fit covariance matrix and rescaled by $\sqrt{\chi^2_\mathrm{red}}$, where $\chi^2_\mathrm{red}$ is the reduced chi-square. This accounts for modest model mismatch and residual excess variance in the image cutouts, yielding centroiding errors that more realistically reflect the residual scatter. The thermal-noise contribution to the astrometric uncertainty is also expected to scale approximately as $\sigma_{\theta,\mathrm{therm}} \approx \theta_{\mathrm{beam}}/(2\,\mathrm{SNR})$, where $\theta_{\mathrm{beam}}$ is the synthesized beam full width at half maximum (FWHM) \citep[e.g.,][]{1988ApJ...330..809R,2020A&ARv..28....6R}, corresponding here to additional localization uncertainties of order $\sim$1--2\,arcmin.

\begin{table}[htbp]
  \centering
  \setlength{\tabcolsep}{2.8pt} 
  \caption{Parameters of the surviving \emph{Stage~II} candidates. 
  Coordinates are from Gaussian centroid fits. Positional uncertainties 
  are 1$\sigma$ errors in arcminutes. All surviving candidates were detected with an effective width of one channel, corresponding to approximately 10\,Hz.}
  \label{tab:stage2_candidates}
  \begin{tabular}{cccccc}
    \hline
    \hline
    Frequency & RA & $\sigma_{\mathrm{RA}}$ & Dec & $\sigma_{\mathrm{Dec}}$ & SNR \\
    (MHz) & (J2000) & (arcmin) & (J2000) & (arcmin) &  \\
    \hline
    60.333372 & 10:18:19.0 & 1.09 & +40:40:37.2 & 1.07 & 15.77 \\
    60.333352 & 10:26:33.6 & 1.47 & +31:23:02.4 & 1.21 & 15.90 \\
    84.333912 & 13:57:33.8 & 1.61 & +16:46:33.6 & 1.25 & 11.47 \\
    \hline
  \end{tabular}
\end{table}

\subsection{Candidate Vetting and Temporal Follow-up}\label{sec:cand_analysis}

\begin{figure*}[htb]
  \centering
  \includegraphics[width=0.9\textwidth]{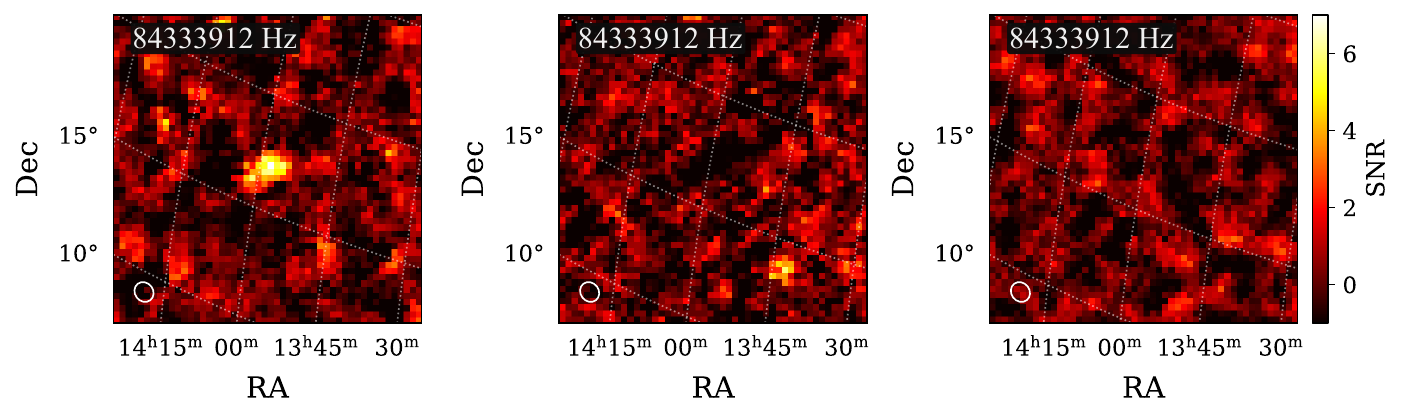}
  \caption{Images of the 84\,MHz candidate in subsequent 30\,s integrations following the original detection shown in \autoref{fig:all_candidates}. The left panel shows the immediately following 30\,s integration, the middle panel the next one after that, and the right panel the final 30\,s integration shown here. The candidate is detected only in the first subsequent image and is absent in the later time steps. The ellipse in the lower-left corner shows the synthesized beam FWHM.}
  \label{fig:one_cand}
\end{figure*}

\subsubsection{Checks Against Local and Orbital Emitters}\label{sec:contaminants}

To assess whether the candidates could arise from a nearby (near-field) emitter such as an aircraft, we repeated the imaging while restricting the \emph{uv} coverage to different baseline ranges.  At our observing frequencies, a source at range $R$ departs from the plane-wave (far-field) approximation when $R \lesssim B^{2}/\lambda$ (Fraunhofer criterion, e.g., \citealt{2017isra.book.....T}), where $B$ is the baseline length and $\lambda$ is the observing wavelength; thus a local emitter is expected to lose coherence on sufficiently long baselines.  We therefore re-imaged the fine channels containing the candidates with a sequence of baseline cuts and measured both the candidates and a bright astrophysical control source (Vir~A in our case) at fixed sky positions.  The candidates persist under baseline selection and show no evidence for additional suppression on longer baselines beyond that expected from reduced \emph{uv} coverage and a changing point-spread function.  We therefore find the signals to be consistent with a far-field source within the sensitivity of these tests.

We cross-matched the candidate sky positions against predicted satellite positions computed from Two-Line Element (TLE) sets obtained via the Space-Track GP\_History archive\footnote{ \url{https://www.space-track.org/documentation} } over 2024 April 5--7. For each candidate, we propagated the queried TLEs to the observation time using the \texttt{skyfield} package \citep{2019ascl.soft07024R} with the SGP4 general-perturbations orbit model \citep{Vallado2006RevisitingSpacetrack3} and compared the resulting topocentric positions (as viewed from the observatory) to the measured candidate coordinates. The queried TLE sets were chosen to cover the major operational satellite constellations and representative navigation, geosynchronous, and debris populations. Using a conservative positional matching criterion based on the candidates’ localization uncertainties and requiring the satellite to be above the local horizon, we find no satellites within the matching radius for any candidate. Moreover, even allowing for additional low-frequency astrometric wander due to ionospheric refraction at these frequencies, we find no clearly compelling satellite association for any candidate at the relevant epoch.

\subsubsection{Temporal Persistence in Adjacent Snapshots}\label{sec:persist}
To further investigate the three surviving \emph{Stage~II} candidates (\autoref{sec:cands}), we form images at their detected frequencies over the next few minutes using the same time integration and frequency range to determine whether the source remains at the same sky position.

Assuming a steady or slowly varying narrowband transmitter, one would generally expect repeated detections in adjacent integrations, possibly with small frequency shifts due to drift. We find that none of the candidates show such persistent behavior in subsequent time steps, except for one event at 84~MHz. For this final candidate, emission is detected again in the following 30~s integration, but not in any of the subsequent time steps (seven integrations in total, spanning approximately three minutes). Images corresponding to some of these subsequent time steps are shown in \autoref{fig:one_cand}.

For the second 84~MHz detection, Gaussian centroiding gives $\mathrm{RA} = 13^{\mathrm h}58^{\mathrm m}11.5^{\mathrm s}$ and $\mathrm{Dec} = +16^\circ52^\prime08.4^{\prime\prime}$, with positional uncertainties of $2.07\arcmin$ and $1.95\arcmin$ in right ascension and declination, respectively, and a peak SNR of about~7.4. Comparing this position to that of the original 84~MHz detection in \autoref{tab:stage2_candidates}, $\mathrm{RA} = 13^{\mathrm h}57^{\mathrm m}33.8^{\mathrm s}$ and $\mathrm{Dec} = +16^\circ46^\prime33.6^{\prime\prime}$, with corresponding uncertainties of $1.61\arcmin$ and $1.25\arcmin$, yields an angular separation of $10.60\arcmin$. Combining the centroiding uncertainties of the two detections in quadrature gives an effective $1\sigma$ positional uncertainty of $3.50\arcmin$, so the two centroids differ at the $\sim3.0\sigma$ level under a stationary-source hypothesis.

As an empirical astrometric check, we applied the same Gaussian-centroiding procedure to Vir~A in the relevant narrowband images. In all candidate images examined, the recovered Vir~A centroid agreed with its catalog position to within approximately \(1\arcmin\). At the time of observation, Vir~A and the 84~MHz candidate were at broadly similar elevations, so they sampled comparable, though not identical, ionospheric conditions. The \(\sim10.60\arcmin\) separation between the two candidate detections is nevertheless much larger than the astrometric offsets seen for Vir~A. Taken together with the higher-resolution behavior discussed in \autoref{sec:high-res}, this positional offset disfavors an astrophysical interpretation and is more consistent with terrestrial or near-Earth interference.

\subsection{Higher-Resolution Analysis}
\label{sec:high-res}

To investigate the nature of the candidates, we reprocessed the raw voltage dumps covering the detection epoch and subsequent intervals at finer temporal and spectral resolution. We generated a sequence of snapshot images with shorter integrations and finer channelization to resolve the signals’ kinematic and morphological behavior.

\begin{figure}[htbp]
  \centering
  \includegraphics[width=\columnwidth]{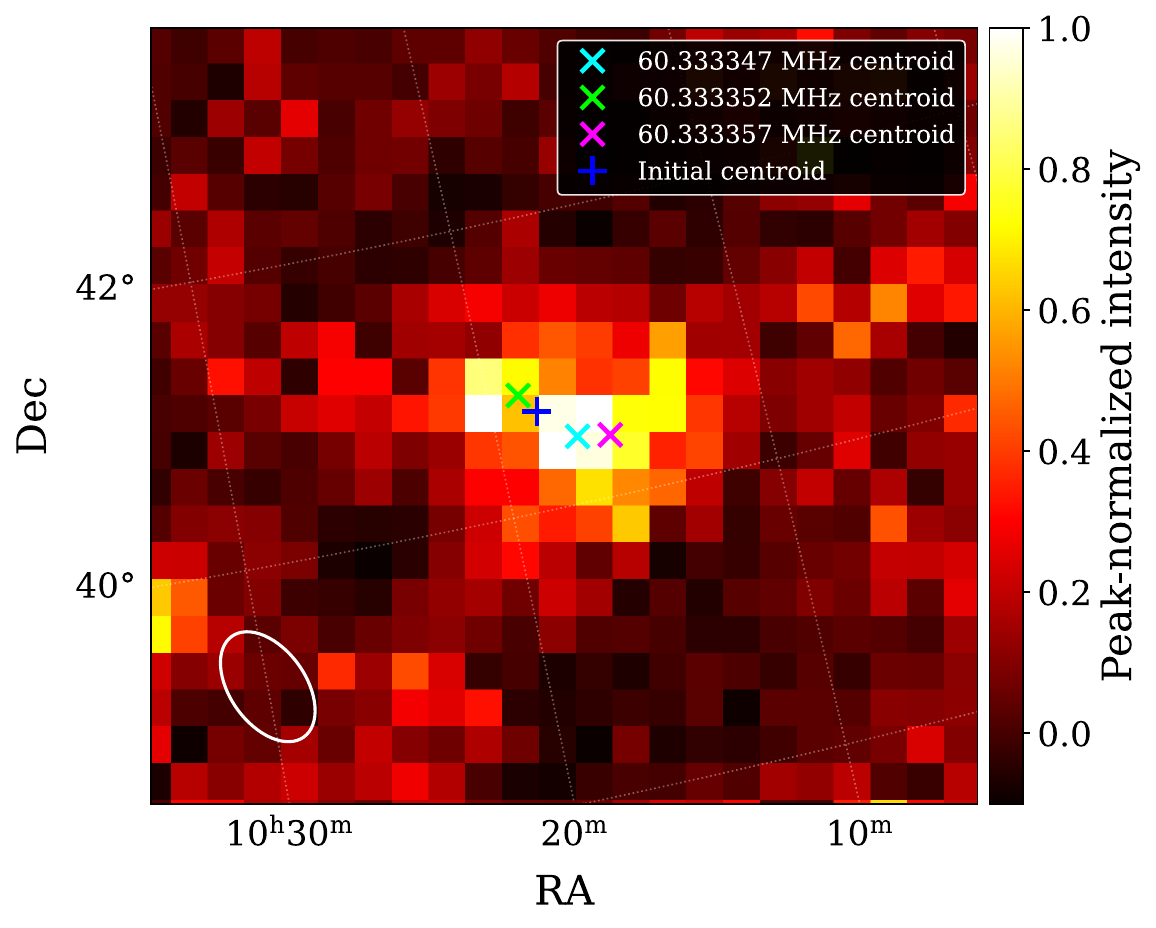}
  \caption{Higher-resolution composite view of one of the 60.33\,MHz candidates re-imaged with 30\,s integration in three adjacent 5\,Hz channels. The background image shows a peak-normalized maximum-composite of the three fine-channel images. Colored crosses show Gaussian-fit centroids measured in the neighboring fine channels, while the blue cross marks the centroid from the original 10\,Hz detection image. The ellipse in the lower-left corner shows the synthesized beam FWHM.}
  \label{fig:60_centroids}
\end{figure}

Reimaging at finer spectral resolution with 30\,s integrations and $\sim5$\,Hz channels shows that both 60.33\,MHz candidates are inconsistent with compact celestial narrowband emitters. In one candidate, shown in \autoref{fig:60_centroids}, the emission appears in multiple adjacent 5\,Hz channels with centroid positions that are not spatially coincident from channel to channel; the second candidate shows similar behavior. The measured channel-to-channel separations exceed \(5\arcmin\) for the lower-frequency candidate and \(10\arcmin\) for the higher-frequency candidate. A true narrowband point source would remain spatially coincident across adjacent fine channels; instead, these coarse-channel detections are blends of spatially inconsistent fine-channel structure rather than single compact sources. As a control, Vir~A is recovered to within \(\sim1\)--\(1.5\arcmin\) in the same images, much smaller than the candidate offsets. We therefore classify both 60.33\,MHz candidates as interference.

\begin{figure}[htbp]
  \centering
  \includegraphics[scale=0.5]{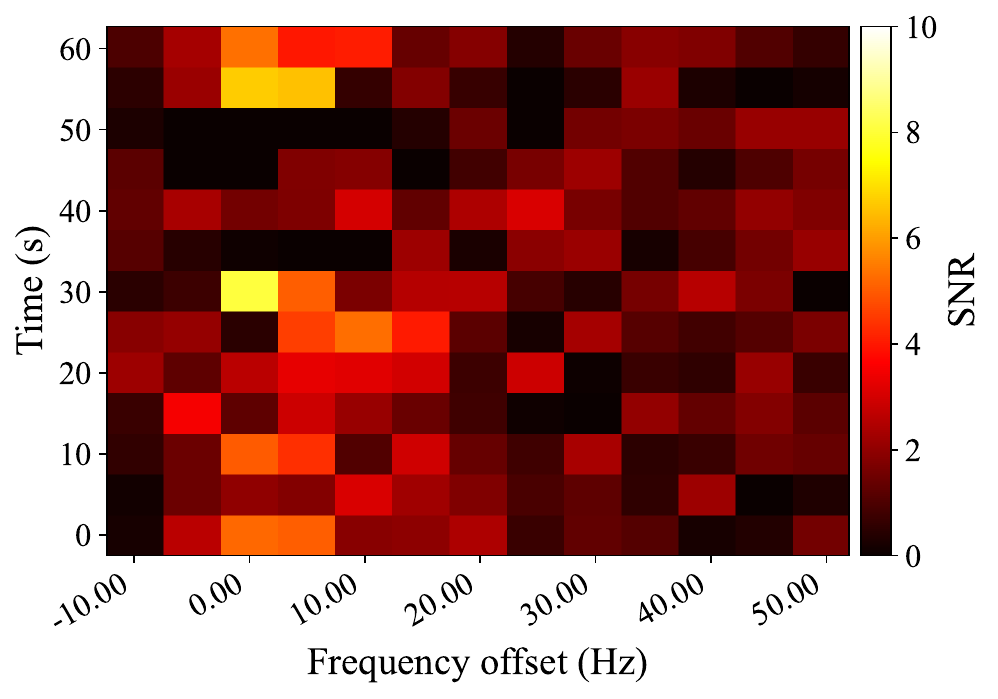}
  \caption{Higher-resolution time--frequency diagram of the 84.333912\,MHz candidate generated with $\simeq5$\,s time resolution and $\simeq5$\,Hz frequency resolution. The horizontal axis shows frequency offset relative to the original detection frequency, and the vertical axis shows time offset from the start of the re-imaged interval.}
  \label{fig:84_dynspec}
\end{figure}

For the remaining candidate at 84.33\,MHz, the temporal follow-up in \autoref{sec:persist} already showed that the signal is inconsistent with a single stationary point source: it reappears only once in the immediately following 30\,s snapshot, and the two detections are offset by $10.60\arcmin$. The higher-resolution reimaging further clarifies its phenomenology (\autoref{fig:84_dynspec}). The dynamic spectrum was generated with $\simeq5$\,s time resolution and $\simeq5$\,Hz frequency resolution. In the time--frequency plane, the emission initially shifts by about three 5\,Hz channels over the first $\sim25$\,s, corresponding to an average drift rate of order $0.6\,\mathrm{Hz\,s^{-1}}$. However, this behavior is not maintained: the signal subsequently reverses or fluctuates between adjacent channels rather than continuing along a coherent monotonic drift track. The emission also appears intermittently on timescales of $\sim10$--15\,s, with brief peaks reaching SNR $\sim8$. This combination of positional inconsistency, intermittent appearance, and irregular fine-scale spectral behavior further supports a terrestrial-RFI interpretation rather than an extraterrestrial emitter, consistent with the need for detailed verification of narrowband technosignature candidates \citep{2021NatAs...5.1153S}.

\section{Results and Constraints}\label{sec:analysis}
\subsection{Technosignature Limits}\label{sec:seti-constraints}

We translate our per-channel image noise into upper limits on transmitter strength using the equivalent isotropic radiated power (EIRP). 
For an unresolved emitter at distance $d$, assuming an intrinsically narrow line with full width $\Delta\nu$ equal to our channel resolution of 10\,Hz, 
the corresponding limit is
\[
\mathrm{EIRP}_{\lim} 
= 4\pi d^{2}\, n_\sigma\, \sigma_\nu\,\Delta\nu
\]

\begin{equation}
\approx n_\sigma\, 
\left(\frac{\sigma_\nu}{100~\mathrm{Jy}}\right)
\left(\frac{d}{1~\mathrm{pc}}\right)^{2} 
1.2\times10^{11}~\mathrm{W}.
\end{equation}

For our 30\,s, $\Delta\nu\simeq10$\,Hz images, we measure a representative per-channel RMS sensitivity of $\sim100$\,Jy  (Sec.~\ref{sec:calib}). Adopting a $10\sigma$ detection threshold, the corresponding upper limits on isotropic transmitter power are $\mathrm{EIRP}_{\lim}\!\approx\!1.2\times10^{14}$\,W for a source at 10\,pc, $1.2\times10^{16}$\,W at 100\,pc, and $1.2\times10^{18}$\,W at 1\,kpc.

An Arecibo-class planetary radar ($\mathrm{EIRP}\!\sim\!2\times10^{13}~\mathrm{W}$) would be detectable in these data out to $\sim3$--$4$~pc for a truly sub-channel ($\lesssim10$~Hz) signal. However, Arecibo operated at S-band ($\sim2.3$~GHz), far above the 50--86\,MHz range considered here. As more frequency-appropriate fiducial benchmarks, one may instead consider powerful VHF/UHF radar-like transmitters, such as surveillance or early-warning systems, with $\mathrm{EIRP}\sim10^{8}$--$10^{9}$~W, corresponding approximately to a $\sim$1\,MW transmitter with antenna gain of order 20--30\,dBi. Such emitters would be detectable only at outer-Solar-System scales, well short of the nearest star, implying that substantially more luminous low-frequency transmitters would be required for interstellar detectability in this survey. 

Although modern targeted surveys (e.g., Breakthrough Listen with GBT/MeerKAT) are typically sensitive enough to detect Arecibo-class transmitters out to $\sim$10--50~pc, our all-sky imaging is less sensitive but provides unprecedented breadth: it is agnostic to sky position and, in principle, probes millions of potential targets at once. Concretely, the 10~pc volume contains $\sim\!300$ stellar systems (nearly complete; \citet{2018AJ....155..265H}), the 100~pc sphere contains a few$\times10^5$ stars (the Gaia 100~pc catalogue lists $\sim3.3\times10^5$ objects) \citep[e.g.,][]{2021A&A...649A...6G,2021A&A...649A...1G}, and a homogeneous local-density extrapolation implies of order $10^8$ stars within 1~kpc. Most stars host planets \citep[e.g.,][]{2012Natur.481..167C,2013PNAS..11019273P,2021A&A...649A...1G}---on average $\sim$one planet per star, with $\gtrsim$\,tens-of-percent occurrence for small planets around Sun-like stars---so wide-field surveys naturally encompass vast numbers of potential planetary systems.

Future, deeper integrations are therefore well justified for persistent or high-duty-cycle narrowband emitters. With $\sim100\times$ longer integrations ($\sim1$~hr), the thermal sensitivity improves by 10, lowering EIRP limits by an order of magnitude for such signals; for intrinsically short-lived emitters, however, longer integrations would dilute the signal instead. We note that Doppler drift/acceleration effects would also become non-negligible and should be taken into account accordingly.

\subsection{Comparison to Other SETI Searches}\label{sec:seti-comparison}

Interpreting EIRP limits across radio technosignature surveys requires careful attention to the assumed signal bandwidth, integration time, detection threshold, and—most importantly—the reference distance at which limits are quoted. For thermal-noise–limited narrowband searches, the limiting EIRP approximately scales as
$\mathrm{EIRP}_{\lim}\propto n_\sigma\,d^{2}\sqrt{\Delta\nu/t}$,
since the per-channel RMS noise satisfies $\sigma_\nu\propto(\Delta\nu t)^{-1/2}$.
As a result, differences of orders of magnitude in EIRP do not necessarily imply comparable differences in raw sensitivity.

Published low-frequency radio technosignature searches remain rare. Previous modern efforts have concentrated mainly near $\sim$100--190\,MHz, including MWA imaging surveys at 103--133\,MHz, 99--122\,MHz, and 98--128\,MHz \citep{2016ApJ...827L..22T,2018ApJ...856...31T,2020PASA...37...35T,2024ApJ...972...76T}, as well as a dual-station LOFAR search at 110--190\,MHz \citep{2023AJ....166..193J}. One published technosignature search that extends substantially below 100\,MHz is the MWA search for narrowband signals from 1I/`Oumuamua at 72--102\,MHz \citep{2018ApJ...857...11T}. In this context, our OVRO--LWA search is one of very few published radio technosignature searches in this frequency regime, extending the literature to 50--86\,MHz with all-sky imaging at $\sim$10\,Hz spectral resolution. This band also overlaps the lower end of the 80--300\,MHz interval highlighted by \citet{2007JCAP...01..020L} as a part of the spectrum where terrestrial radio leakage could be strong.

At GHz frequencies, targeted surveys achieve the deepest EIRP limits for nearby stars. The Breakthrough Listen survey with the Green Bank Telescope reports observations of 1327 main-sequence stars within 50~pc using 300~s pointings, a spectral resolution of 2.79~Hz, and a $10\sigma$ detection threshold, yielding a minimum detectable EIRP of $2.1\times10^{12}$~W at 50~pc \citep{2020AJ....159...86P}. Earlier GBT Breakthrough Listen observations of 692 nearby stars at 1.1--1.9~GHz report EIRP limits of order $10^{13}$~W for comparable distances \citep{2017ApJ...849..104E}. These surveys benefit from low system temperatures, Hz-scale channelization well matched to intrinsically narrow transmitters, and targeted integrations on nearby stellar systems.

Commensal GHz surveys trade per-target depth for survey breadth. The COSMIC search using VLASS data analyzes short ($\sim$8~s) beamformed segments with 7.63~Hz spectral resolution and an $8\sigma$ detection threshold, reporting EIRP limits ranging from $2.3\times10^{11}$~W to $2\times10^{16}$~W for sources spanning distances from 4.3~pc to $\sim$1.3~kpc \citep{2025AJ....169..122T}. 

At low radio frequencies, wide-field surveys provide complementary constraints over very large target populations. A dual-station LOFAR search at 110--190~MHz employs $\sim$3~Hz spectral resolution, 15~minute integrations, and a $10\sigma$ threshold, and excludes continuous narrowband transmitters with EIRP $\gtrsim10^{17}$~W for $\sim4\times10^{5}$ stellar systems with a mean distance of $\sim1.3$~kpc \citep{2023AJ....166..193J}. 

MWA imaging-based searches emphasize extremely wide instantaneous sky coverage at the cost of coarse spectral resolution. Several MWA surveys at 100--130~MHz use 10~kHz channels and multi-hour integrations, reporting best-case EIRP limits of $\sim10^{13}$--$10^{14}$~W for the nearest stellar systems \citep[e.g.,][]{2016ApJ...827L..22T,2018ApJ...856...31T,2020PASA...37...35T,2022PASA...39....8T}. However, these works explicitly note that a truly narrow ($\sim$10~Hz) transmitter would be significantly diluted in a 10~kHz channel, and that the corresponding EIRP limits would increase by a factor $\sim10^3$ for such signals. Consequently, coarse-channel imaging surveys are intrinsically less sensitive to sub-Hz or few-Hz technosignatures unless reprocessed at finer spectral resolution.

In this context, our limits are naturally intermediate between deep, targeted GHz surveys and low-frequency, coarse-resolution imaging searches. Although our integrations are short (30~s) and the low-frequency sky noise is high, our 10~Hz channelization is well matched to narrowband transmitters and avoids the bandwidth-dilution penalty. When compared at similar distances and for comparable signal bandwidth assumptions, our EIRP limits are therefore consistent with expectations from radiometer scaling and complement existing surveys by providing uniform, all-sky coverage at tens of MHz.

\subsection{Axion-Like Particle Prospects}\label{sec:axion}

The expected flux of photons converted from axions in the neutron-star magnetosphere depends on the field geometry, plasma profile, source distance, and the axion–photon coupling $g_{a\gamma\gamma}$, with $S_{\rm conv}\!\propto\! g_{a\gamma\gamma}^2\rho_{\rm DM}/D^2$, where $S_{\rm conv}$ is the radio flux density from axion–photon conversion, $\rho_{\rm DM}$ is the local dark-matter density, and $D$ is the source distance. Benchmark calculations in the literature indicate sub-mJy to mJy-scale fluxes in favorable cases at $g_{a\gamma\gamma}\sim10^{-12}\,\mathrm{GeV}^{-1}$, though predictions span orders of magnitude. With our representative per-channel RMS of $\sim\!100$\,Jy and a conservative $10\sigma$ threshold, single-epoch limits are far above existing astrophysical bounds. Progress could come from deeper pointed integrations and stacking ensembles of neutron stars \citep{2018PhRvL.121x1102H,2021JCAP...11..013M,2021JHEP...09..105B}. In addition to Galactic pulsars, globular clusters may provide attractive intermediate targets because they host rich neutron-star populations and, in some cases, dozens of known pulsars \citep{2019PhRvD..99l3021S}. For instance, increasing the collecting area through a future low-frequency dipole array---for example, a possible extension of DSA to lower frequencies or a comparable instrument---together with stacking all currently known Galactic pulsars (several thousands), would yield a combined sensitivity gain of hundreds of times. Bridging the remaining gap to reach mJy-level theoretical benchmarks would then require extending integration times from seconds to order $\sim\!10^3$ hours.

At the same time, stimulated $a\!\to\!\gamma\gamma$ emission scales with $g_{a\gamma\gamma}^2\,\rho_{\rm DM}\,T_b$, where $T_b$ is the local radio brightness temperature, making the Galactic Center---where $T_b$ reaches $\sim10^5$\,K at 70\,MHz---the optimal target. However, strong inner-Galaxy propagation effects, including free--free absorption and scattering, may partly offset this advantage at tens of MHz. Nevertheless, our snapshot flux limits imply $g_{a\gamma\gamma}$ constraints that remain many orders of magnitude weaker than those from state-of-the-art haloscope experiments. Increasing the integration time would improve sensitivity, but a larger effective collecting area (e.g., by coherently combining more dipoles) is even more beneficial and could achieve substantially deeper limits. Such improvements would not only enhance ALP searches but also strengthen the array’s capabilities for technosignature and narrowband SETI experiments.

We note that coherent stacking of ultra-narrowband images requires placing all of them on a common astrometric/phase screen. At tens of MHz, direction-dependent ionospheric refraction produces arcminute-level snapshot offsets that vary across the field. In practice, one could (i) estimate an image-plane refractive field from bright compact sources and de-warp each channel/epoch accordingly; or (ii) solve for a total electron content (TEC) phase screen via direction-dependent calibration, then reimage with those solutions applied \citep[e.g.,][]{2009A&A...501.1185I,2016ApJS..223....2V}. Another alternative is to apply an \emph{uv} taper and restrict to ionospherically stable spans before stacking. Application of these strategies is beyond the scope of this paper and could be pursued in future studies.

\section{Conclusions and Future Steps}\label{sec:summary}
We have demonstrated an imaging-domain technosignature search across 50--86\,MHz with 10\,Hz spectral resolution and 30\,s integrations, forming over $3\times10^{6}$ all-sky images for a single epoch. More broadly, this work extends published technosignature searches into a still sparsely explored sub-100\,MHz regime, while combining all-sky imaging with ultra-fine spectral resolution. We examined propagation and kinematic effects relevant to ultra-narrowband tones in this regime, finding that most plausible signals are not significantly spectrally broadened or drifted relative to our 10\,Hz channels over 30\,s integrations.

Candidate identification proceeded in two stages. \emph{Stage~I} applied matched filtering and false-discovery-rate control to the standardized image cubes, yielding an initial candidate set. \emph{Stage~II} removed artifacts and extended or corrupted images through automated quality cuts and visual inspection, leaving three candidates above $10\sigma$. To determine the physical nature of these surviving candidates, we reprocessed the raw voltage data at finer temporal and spectral resolution. This analysis revealed that the two 60\,MHz candidates are spectrally and spatially inconsistent with compact celestial narrowband emitters and are best explained as interference, while the 84\,MHz candidate showed intermittent emission with a significant positional shift between detections, consistent with terrestrial interference.

The representative per-channel sensitivity of around $100$\,Jy implies $10\sigma$ EIRP limits of $\sim10^{14}$\,W at 10\,pc and $\sim10^{18}$\,W at 1\,kpc for unresolved, non-drifting lines. Although this sensitivity is lower than that of targeted GHz-band surveys, our approach offers uniform sky coverage, simultaneously monitoring millions of stellar systems without selection bias. Future work will leverage longer integrations to perform explicit drift-rate searches and analyses of specific source classes, such as ALP conversions in neutron-star magnetospheres.

This analysis was carried out using a conservative data-handling configuration in which input/output (I/O) operations were performed directly from the shared Lustre parallel filesystem, without staging data onto node-local Non-Volatile Memory Express (NVMe) solid-state drives. In this setup, the dominant bottleneck is data transfer rather than cross-correlation or imaging itself. Using node-local NVMe storage for intermediate data products, together with reusing calibration solutions across adjacent time steps, improves the end-to-end throughput of the pipeline by nearly an order of magnitude. These optimizations will enable deeper integrations, analysis of multiple combined epochs, and extension of the present search to a larger fraction of the OVRO--LWA observing band.

This approach can also be used to study the spectral structure of terrestrial and orbital RFI sources in unprecedented detail. The 10\,Hz spectral resolution enables analysis of persistent narrow carriers from broadcast and navigation transmitters, low-Earth-orbit satellite beacons, and other stable narrowband emitters. These observations could provide a valuable diagnostic tool for characterizing RFI morphology, identifying emission mechanisms, and improving excision strategies in future low-frequency interferometric surveys.

\begin{acknowledgments}
We thank the anonymous referee and Stella Koch Ocker for useful comments on the manuscript. This material is based in part upon work supported by the National Science Foundation under grant numbers AST-1828784 and AST-2206574, the Simons Foundation (668346, JPG), the Wilf Family Foundation and Mt. Cuba Astronomical Foundation. This research is also enabled by development funded by Schmidt Sciences for the Deep Synoptic Array project (DSA). The DSA is part of the Eric and Wendy Schmidt Observatory System in partnership with the California Institute of Technology. N.K. acknowledges support from the Schmidt Academy for Software Engineering at the California Institute of Technology.
\end{acknowledgments}

\bibliography{seti_axion}{}
\bibliographystyle{aasjournalv7}



\end{document}